\documentstyle[amssymb,aps]{revtex}
\begin{document}
\draft
\title{Note on the stability of a confined Bose gas: Comment on quant-ph/9712030 by
L. Salasnich}
\author{F. Brosens\thanks{%
Senior Research Associate of the FWO (Fonds voor Wetenschappelijk Onderzoek
- Vlaanderen).}, J. T. Devreese\thanks{%
Also at Universiteit Antwerpen (RUCA) and Technische Hogeschool Eindhoven,
Nederland.}}
\address{Departement Natuurkunde, Universiteit Antwerpen (UIA),\\
Universiteitsplein 1, B-2610, Antwerpen }
\author{L. F. Lemmens}
\address{Departement Natuurkunde, Universiteit Antwerpen (RUCA), Groenenborgerlaan\\
171, B-2020, Antwerpen }
\date{December 19, 1997}
\maketitle
\pacs{PACS Numbers: 03.75.Fi, 05.30.Jp}

\begin{abstract}
We demonstrate analytically that a Bose-Einstein condensate confined in a
harmonic trap with zero-range attractive interparticle interactions is
unstable if there is more than 1 boson. Replacing the zero-range interaction
by a short-range attractive interaction lifts the instability, and leads to
a pronounced clustering, by which the particles leak out of the condensate.
\end{abstract}

In a recent preprint\cite{Salasnich}, Salasnich claims to confirm
analytically the numerical calculations of Edwards and Burnett\cite{Edwards}
and of Dalfovo and Stringari\cite{Dalfovo} on the ground state stability of
the Bose-Einstein condensate (BEC). Also the maximum number of bosons is put
forward, for which the BEC with attractive zero-range interaction should be
stable. However, these conclusions are based on an incomplete analysis of
his upper bound $K$ to the ground state energy (equation (17) in Ref. \cite
{Salasnich}), which reads (in the units $\hbar =1$ and $m=1$, which are used
throughout this paper) 
\begin{equation}
K=\frac{3}{4}N\frac{1}{\sigma ^{2}}+\frac{3}{4}N\Omega ^{2}\sigma ^{2}+N^{2}%
\frac{B}{\left( 2\pi \right) ^{3/2}}\frac{1}{\sigma ^{3}}
\end{equation}
$N$ is the number of bosons, $\Omega $ is the frequency of the confinement
potential (assumed isotropic), $B$ is the strength of the zero-range
interparticle interaction, and the variational parameter $\sigma >0$ is the
standard deviation in a Gaussian trial wave function.

From a mathematical point of view, one can hardly overlook the minimum $%
K\rightarrow -\infty $ for $\sigma \rightarrow 0$ in the case $B<0$, i.e.
for an attractive two-particle interaction. The fact that this instability
also seems to occur for $N=1$ is an artefact of an approximation made in the
derivation of $K$. One readily checks that $N^{2}$ should be replaced by $%
N\left( N-1\right) .$

The instability with mean distance $\sigma \rightarrow 0$ signifies the
collapse of the wave function, and the maximum number $N^{\max }$ of
particles, discussed in\cite{Salasnich}, should then not be interpreted as
the number above which the BEC collapses, but as the number of bosons below
which a local minimum exists in the Gross-Pitaevskii\cite{Gross-Pitaevskii}
functional. This poses the question: how long does it take before the bosons
in this metastable state tunnel over the energy maximum into the collapsed
state, thereby leaking out of the condensate?

One might object that this singularity in the Gross-Pitaevskii functional in
the attractive case is a consequence of considering a zero-range
interaction, and one might hope that the instability would disappear for a
more realistic interparticle interaction. However, the Gross-Pitaevskii
approach as it stands is hardly applies in this situation.

Nevertheless, some progress can be made, using the path integral approach
which we developed earlier\cite{BDL1,BDL2}, and which allows to determine
the thermodynamical quantities, the density and the pair correlation
function of a harmonic model system.

By applying the Jensen-Feynman inequality with the trial action resulting
from this harmonic model system, one finds the following upper bound $E_{v}$
to the ground state energy: 
\begin{equation}
E_{v}=\frac{3}{4}N\frac{w^{2}+\Omega ^{2}}{w}-\frac{3}{4}\frac{\left(
w-\Omega \right) ^{2}}{w}+\frac{1}{2}N\left( N-1\right) \int d^{3}{\bf r}%
v_{2}\left( {\bf r}\right) g\left( {\bf r}\right)
\end{equation}
where $w$ is a variational parameter, $v_{2}\left( {\bf r}\right) $ is the
interparticle interaction potential, and $g\left( {\bf r}\right) $ is the
pair correlation function of the harmonic model system. The variational
parameter $w$ describes the frequency of the $3\left( N-1\right) $ internal
degrees of freedom of the harmonic model system, and takes over the role of
the parameter $1/\sigma ^{2}$ in \cite{Salasnich}.

In the zero-temperature limit the pair correlation function $g\left( {\bf r}%
\right) $ as derived in \cite{BDL1,BDL2} is given by 
\begin{equation}
g\left( {\bf r}\right) =\left( \frac{w}{2\pi }\right) ^{3/2}e^{-\frac{1}{2}%
wr^{2}}
\end{equation}
The upper bound $E_{v}$ to the ground state energy can be minimized with
respect to $w$ for a rather large class of two-particle interactions. In the
zero-range case, with $v_{2}\left( {\bf r}_{i}-{\bf r}_{j}\right) =-\left|
B\right| \delta \left( {\bf r}_{i}-{\bf r}_{j}\right) $, our approach leads
to the conclusion that $E_{v}\rightarrow -\infty ,$ corresponding to the
collapse of the wave function. However, this singularity disappears if a
finite-range interaction potential is introduced. For instance, a
single-step potential of the form 
\begin{equation}
v_{2}\left( {\bf r}\right) =-\left| V\right| \Theta \left( R-r\right)
\end{equation}
with the range $R$ in the order of a few Bohr radii requires a depth of the
order $V\sim 10^{8}\Omega $ if a scattering length $a=-14.5$\AA\ is assumed 
\cite{Hulet}. Preliminary results reveal that in this case a {\em local}
minimum is found for $w\sim \Omega $ if $N\lesssim 2875$, but that a {\em %
global} minimum for $E_{v}$ is found for $w$ near $10^{8}\Omega $, separated
from the local minimum by a huge energy barrier of order $V.$

A fully documented analysis of the (repulsive) Rb case, including anisotropy
and temperature dependence will be presented in a forthcoming paper\cite
{Tempere}. For the $^{7}Li$ case a detailed study is in preparation. But
even without the numerical details for the specific alkali metals of
interest, the present analysis allows for a few analytical conclusions.

In the case of a zero-range attractive interaction between the bosons in a
harmonic confining potential, the system is unstable for $N\geq 2$, and the
ground state wave function collapses. The minimum in the Gross-Pitaevskii
functional, found in\cite{Salasnich}, is a {\em local} minimum describing
the BEC as a metastable state. Our generalization for a finite-range
interaction, based on the harmonic model system \cite{BDL1,BDL2} shows that
this instability of the ground state energy is an artefact of the $\delta $
function-like interparticle interaction. For more realistic interparticle
interactions, this singularity in the energy is replaced by a pronounced
minimum, corresponding to an extreme narrowing of the wave function, with a
width of the order of the range of the attractive interaction potential.

\end{document}